\documentclass[aps,prb,showpacs,twocolumn,floats,epsfig]{revtex4}
\usepackage{amssymb}
\usepackage{amsbsy}
\usepackage{amsmath}
\usepackage{epsfig}
\newcommand\beq{\begin{equation}}
\newcommand\eeq{\end{equation}}
\newcommand\bea{\begin{eqnarray}}
\newcommand\eea{\end{eqnarray}}
\newcommand\al{\alpha}
\newcommand\De{\Delta}
\newcommand\ep{\epsilon}
\newcommand\ga{\gamma}
\newcommand\si{\sigma}
\newcommand\dg{\dagger}
\newcommand\non{\nonumber}

\begin{document}

\title{Magnetotransport of Dirac Fermions on the surface of a topological
insulator}

\author{S. Mondal$^1$, D. Sen$^2$, K. Sengupta$^1$, and R. Shankar$^3$}

\affiliation{$^1$Theoretical Physics Division, Indian Association for
the Cultivation of Sciences, Kolkata 700 032, India \\
$^2$Center for High Energy Physics, Indian Institute of Science,
Bangalore 560 012, India \\
$^3$The Institute of Mathematical Sciences, C.I.T. Campus, Chennai 600 113,
India}

\date{\today}

\begin{abstract}
We study the properties of Dirac fermions on the surface of a
topological insulator in the presence of crossed electric and
magnetic fields. We provide an exact solution to this problem and
demonstrate that, in contrast to their counterparts in graphene,
these Dirac fermions allow relative tuning of the orbital and Zeeman
effects of an applied magnetic field by a crossed electric field
along the surface. We also elaborate and extend our earlier results
on normal metal-magnetic film-normal metal (NMN) and normal
metal-barrier-magnetic film (NBM) junctions of topological
insulators [Phys. Rev. Lett. {\bf 104}, 046403 (2010)]. For NMN
junctions, we show that for Dirac fermions with Fermi velocity
$v_F$, the transport can be controlled using the exchange field
${\mathcal J}$ of a ferromagnetic film over a region of width $d$.
The conductance of such a junction changes from oscillatory to a
monotonically decreasing function of $d$ beyond a critical
${\mathcal J}$ which leads to the possible realization of magnetic
switches using these junctions. For NBM junctions with a potential
barrier of width $d$ and potential $V_0$, we find that beyond a
critical ${\mathcal J}$, the criteria of conductance maxima changes
from $\chi= e V_0 d/\hbar v_F = n \pi$ to $\chi= (n+1/2)\pi$ for
integer $n$. Finally, we compute the subgap tunneling conductance of
a normal metal-magnetic film-superconductor (NMS) junctions on the
surface of a topological insulator and show that the position of the
peaks of the zero-bias tunneling conductance can be tuned using the
magnetization of the ferromagnetic film. We point out that these
phenomena have no analogs in either conventional two-dimensional
materials or Dirac electrons in graphene and suggest experiments to
test our theory.
\end{abstract}


\maketitle

\section{Introduction}

Topological insulators with time reversal symmetry in two and three
dimensions (2D and 3D) have been studied extensively in recent
years, both theoretically and experimentally
\cite{zhang1,hasan1,kane1,kane2,qi1,exp2,expt1,hasan2}. The 3D
topological insulators can be characterized by four integers $\nu_0$
and $\nu_{1,2,3}$ \cite{kane2}. The first integer specifies the
class of topological insulators as strong ($\nu_0=1$) or weak
($\nu_0=0$), while the last three integers characterize the
time-reversal invariant momenta of the system given by $\vec M_0 =
(\nu_1 \vec b_1, \nu_2 \vec b_2, \nu_3 \vec b_3)/2$, where $\vec
b_{1,2,3}$ are reciprocal lattice vectors. The topological features
of strong topological insulators (STI) are robust against the
presence of time-reversal invariant perturbations such as disorder
and lattice imperfections. It has been theoretically predicted
\cite{kane2,zhang1} and experimentally verified \cite{hasan1} that
the surface of a STI has an odd number of Dirac cones whose
positions are determined by the projection of $\vec M_0$ on to the
Brillouin zone of the surface. The position and number of these
cones depend on both the nature of the surface concerned and the
integers $\nu_{1,2,3}$. For compounds such as $\rm HgTe$ and ${\rm
Bi_2 Se_3}$, specific surfaces with a single Dirac cone near the
$\Gamma$ point of the 2D Brillouin zone have been found
\cite{hasan1,exp2,hasan2}. Such a Dirac cone is described by the
Hamiltonian \bea H &=& \int \frac{dk_x dk_y}{(2\pi)^2} ~\psi^\dg
(\vec k) ~(\hbar v_F {\vec \si} \cdot {\vec k} - \mu I) ~\psi(\vec
k), \label{ham1} \eea where $\vec \si (I)$ denotes the Pauli
(identity) matrices in spin space, $\psi= (\psi_{\uparrow},
\psi_{\downarrow})^T$ is the annihilation operator for the Dirac
spinor ($T$ denotes the transpose of a row vector), $v_F$ is the
Fermi velocity, and $\mu$ is the chemical potential \cite{kane4}.
Recently, several novel features of these surface Dirac electrons
such as the existence of Majorana fermions in the presence of a
magnet-superconductor interface on the surface \cite{kane4,been1},
generation of time-reversal symmetric $p_x + i p_y$-wave
superconducting state via proximity to a $s$-wave superconductor
\cite{kane4}, anomalous magnetoresistance of ferromagnet-ferromagnet
junctions \cite{tanaka1}, and novel spin textures with chiral
properties \cite{hasan2} have been studied in detail. Further it has
been shown in Ref.\ \onlinecite{mondal1} that it is possible to
realize a magnetic switch by magnetically tuning the transport of
Dirac fermions with a proximate ferromagnetic film. However, the
response of these fermions in the presence of crossed electric and
magnetic fields has not been studied so far. Another aspect of such
fermions, namely, their transport through a normal metal-magnetic
film-superconductor (NMS) junction has also not been explored.

\begin{figure} \includegraphics[width=\linewidth]{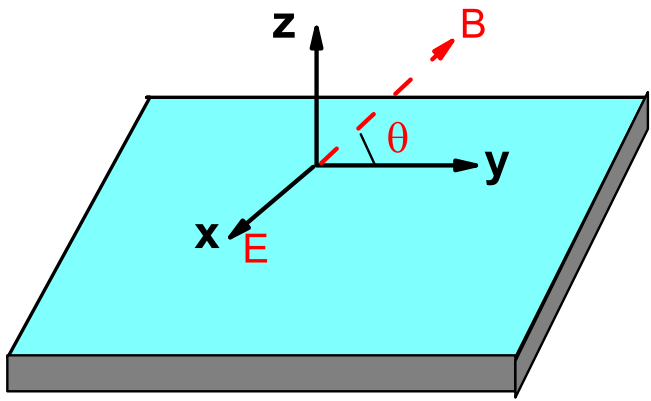}
\caption{Schematic representation of the crossed electric and magnetic field
geometry. The electric field is applied along $x$ while the magnetic field is
in the $y-z$ plane. See text for details.} \label{fig1a} \end{figure}

In this work, we study several magnetotransport properties of these
surface Dirac fermions in experimentally realizable situations. We first
study the properties of the fermions in the
presence of crossed magnetic [$\vec B=(0,B \cos \theta, B \sin
\theta )$] and electric fields [${\vec {\mathcal E}}=({\mathcal
E},0,0)$] as shown in Fig.\ \ref{fig1a}. We present an exact
solution of this problem and show that for $\beta ={\mathcal E}/(v_F
B \sin \theta) \le 1$, the relative contributions of
the Zeeman and the orbital terms to the Landau level energies, and
hence their magnetic field dependence, can be tuned by varying
either the strength of the applied electric field or the tilt of the
applied magnetic field.  We also show that for ${\mathcal E} > v_F B 
\sin \theta$, the conductance of these Dirac fermions has an unconventional dependence
on the tilt angle $\theta$ of the applied magnetic field and that
this dependence can be used to realize electric-field controlled
switching.

\begin{figure} \includegraphics[width=0.8\linewidth]{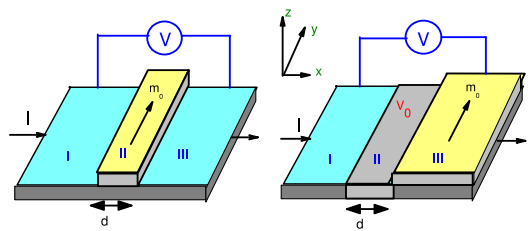}
\caption{Proposed experimental setups. Left panel: Schematic representation
of a NMN junction. The ferromagnetic film extends over region II of width
$d$ providing an exchange field in this region. Right panel: Schematic
representation of a NBM junction. The film extends over region III while
the region II has a barrier characterized by a voltage $V_0$. $V$ and $I$
denote the bias voltage and current across the junctions respectively.
See text for details.} \label{fig2a} \end{figure}

The second study involves an extension of the results obtained in
Ref.\ \onlinecite{mondal1} regarding transport in normal
metal-magnetic film-normal metal (NMN) and normal
metal-barrier-magnetic film (NBM) junctions of topological
insulators. The relevant experimental geometries are shown in Fig.\
\ref{fig2a}. We study the transport of these Dirac electrons across
a region with a width $d$ where there is a proximity-induced
exchange field ${\mathcal J}$ arising from the magnetization $\vec m
= m_0 \hat y$ of a proximate ferromagnetic film as shown in the left
panel of Fig.\ \ref{fig2a}. We demonstrate that the tunneling
conductance $G$ of these Dirac fermions through such a junction can
either be an oscillatory or a monotonically decaying function of the
junction width $d$. One can interpolate between these two
qualitatively different behaviors of $G$ by changing $m_0$ (and thus
${\mathcal J}$) by an applied in-plane magnetic field leading to the
possible use of this junction as a magnetic switch. We also study
the transport properties of Dirac fermions across a barrier
characterized by a width $d$ and a potential $V_0$ in region II with
a magnetic film proximate to region III as shown in the right panel
of Fig.\ \ref{fig2a}. We note that it is well known from the context
of Dirac fermions in graphene \cite{neto1} that such a junction, in
the absence of the induced magnetization, exhibits transmission
resonances with maxima of transmission at $\chi= e V_0 d/\hbar v_F =
n \pi$, where $n$ is an integer. Here we show that beyond a critical
strength of $m_0$, the maxima of the transmission shifts to $\chi=
(n+1/2) \pi$. Upon further increasing $m_0$, one can reach a regime
where the conductance across the junctions vanishes. We also point
out that such NMN and NBM junctions can be used to determine the
exact form of the Dirac Hamiltonian on the surface of the
topological insulator.

\begin{figure} \includegraphics[width=0.8 \linewidth]{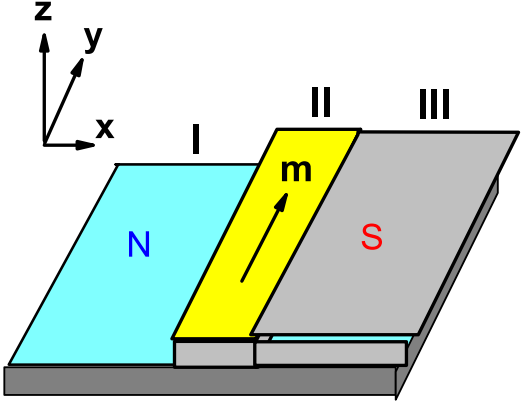}
\caption{Schematic representation of the NMS junction. The ferromagnetic
film is on region II while superconductivity is induced in region III via a
proximate superconducting film. See text for details.} \label{fig3a}
\end{figure}

Finally, we study the transport of Dirac fermions across a NMS
junction as shown in Fig.\ \ref{fig3a}. The intermediate region
(region II) in this junction has a thickness $d$ with a proximate
ferromagnetic film providing a magnetization $M$, while
superconductivity is introduced in region III via the proximity
effect. We provide a detailed analysis of the subgap tunneling
conductance $G$ of these NMS junctions as a function of the applied
voltage $V$ and magnetization $M$. In particular, we point out that
the positions of the maxima of the zero-bias tunneling conductance
in such NMS junctions as a function of the width $d$ of the magnetic
film can be varied by tuning the induced magnetization $M$. We
stress that the properties of the Dirac fermions elucidated in all
these studies are a consequence of their spinor structure in
physical spin space, and thus have no analogs for either
conventional Schr\"odinger electrons in 2D or Dirac electrons in
graphene \cite{kat1,comment1,comment2}.

The organization of the rest of the paper is as follows. In Sec.\
\ref{cmef}, we study the properties of the Dirac fermions in the
presence of crossed electric and magnetic fields. This is followed
by the study of NMN and NBM junctions of these Dirac materials in
Sec.\ \ref{nmjn}. In Sec.\ \ref{supjn}, we study the transport
properties and subgap tunneling conductance of NMS junctions of
topological insulators. Finally we discuss possible experimental
verification of theory and conclude in Sec.\ \ref{conc}.

\section{Crossed electric and magnetic fields}
\label{cmef}

We begin with the properties of Dirac electrons in a crossed
electric and magnetic field as shown in Fig.\ \ref{fig1a}. The
Hamiltonian for the Dirac Fermions for this case can be written as
\bea H &=& \int d^2r \psi^\dg (\vec r) [v_F {\vec \si} \cdot \Pi -
\mu I
- g \mu_B {\vec \si} \cdot {\vec B} - e{\mathcal E} x ] \psi (\vec r), \non \\
\label{emham1} \eea where $\vec \Pi = -i \hbar {\vec \nabla} - e
\vec A$ is the canonical momentum, $c$ is set to unity, $g$ is the
gyromagnetic ratio, $\mu_B$ is the Bohr magneton, and we choose the
vector potential to be $\vec A = (0,Bx \sin \theta, -B x \cos
\theta)$. Note that here the Zeeman term does not determine the spin
quantization axis of the Dirac electrons due to the presence of the
$\vec \Pi$ term. Thus the in-plane component of the magnetic field,
which enters the Hamiltonian only through the Zeeman term, only
provides a constant shift to $k_y$ which can be gauged away. This
property of the Dirac fermions is distinct from their counterpart in
graphene. For ${\mathcal E}=0$, Eq.\ (\ref{emham1}) admits a
straightforward solution and yields the Landau level spectrum
\bea E_n &=& \pm \hbar v_F l_B^{-1} \sqrt{|n| + \al B \sin \theta} \quad
{\rm if } \, n \ne 0 \non \\
&=& -|g \mu_B B \sin \theta| \quad {\rm if} \, n=0, \label{lan1}
\eea where $\al = g^2 \mu_B^2 /\hbar v_F^2 e$, $l_B= \sqrt{\hbar/[eB
\sin \theta]}$ is the magnetic length and we also define $l_B^0 =
l_B (\theta=\pi/2)$ for later use. The $n=0$ state is non-degenerate
as is also known from analogous studies of Landau levels in graphene
\cite{neto1}. For the Dirac electrons on the surface of ${\rm Hg
Te}$, $v_F \simeq 5.5 \times 10^5$ m/s, so that $\al \simeq
10^{-4}/T$ leading to a negligible contribution of the Zeeman term
in the spectrum.

The situation changes when an electric field is applied along $x$.
In this case, for ${\mathcal E} \le v_F B \sin \theta$, one can
define a boost parameter $\beta = {\mathcal E}/[v_F B \sin \theta]
\le 1$ and carry out a Lorentz transformation \cite{vinu1} \bea
x'&=&x, \quad y'= \ga (y-\beta v_F t), \quad t'= \ga(t-\beta y/v_F),
\non \\
{\mathcal E}' &=& \ga( {\mathcal E} - \beta B \sin \theta),
\non \\ B'\sin \theta' &=& \ga( B \sin \theta - \beta {\mathcal E}), \non \\
\psi'({\vec r}^{~'}) &=& \exp [-\si_y {\rm arctanh} (\beta)/2] ~\psi (\vec r),
\label{lorentr} \eea
where $\ga=(1-\beta^2)^{-1/2}$. In the boosted frame the Schr\"odinger
equation reads
\bea E'_n \psi' &=& \Big[ -i \hbar v_F \left( \si_x \partial_x + \si_y
(\partial_{y'} - i \frac{e B' \sin \theta'}{ v_F} x) \right) \non \\
&& ~~- g \mu_B \si_z B \sin \theta \Big] \psi' . \label{sch1} \eea
Note that such a boost transformation affects the orbital part of
the magnetic field only; the Zeeman field remains unchanged. The energy
eigenvalues of Eq.\ (\ref{sch1}) can be easily obtained and are given by
\bea E'_n &=& \pm \hbar v_F l_B^{'\,-1} \sqrt{|n| + \al B \sin \theta}
\quad {\rm if} \, n \ne 0 \non \\
&=& -|g \mu_B B \sin \theta| \quad {\rm if} \, n=0, \eea where
$l_B^{'}=\sqrt{\hbar/[e B' \sin \theta']}$. Then a reverse boost to
the ``laboratory" frame yields \bea E_{n}(k_y) &=& \pm \hbar v_F
l_B^{-1} \ga^{-3/2} \sqrt{|n| + \al B
\ga \sin \theta} \non \\
&& -\beta \hbar v_F k_y \quad {\rm if} \, n \ne 0 \non \\
&=& -\ga^{-1} |g \mu_B B \sin \theta| -\beta \hbar v_F k_y \quad
{\rm if } \, n=0. \label{lan2} \eea

\begin{figure} \includegraphics[width=\linewidth]{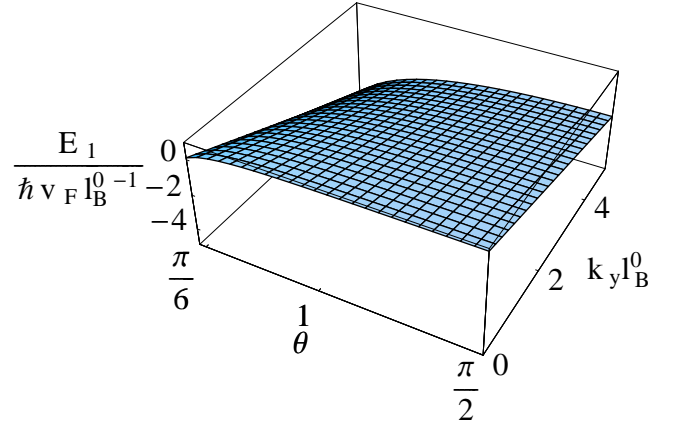}
\caption{Plot of Landau level energy $E_n$ for $n=1$ as a function
of the tilt angle $\theta$ and transverse momentum $k_y$ for a fixed
electric field $\mathcal E=0.5 v_F B$. Note that $\theta=\pi/6$
corresponds to $\beta=1$.} \label{figseciia} \end{figure}

Eq.\ (\ref{lan2}) is one of the central results of this section. It
demonstrates that a collapse of the Landau levels for the Dirac
fermions can be induced by varying either the electric field for a
fixed tilt of the applied magnetic field or by varying the magnetic
field tilt for a fixed electric field. A plot of the energy level as a
function of this tilt and the transverse momentum $k_y$ is shown in
Fig.\ \ref{figseciia}.

\begin{figure} \includegraphics[width=\linewidth]{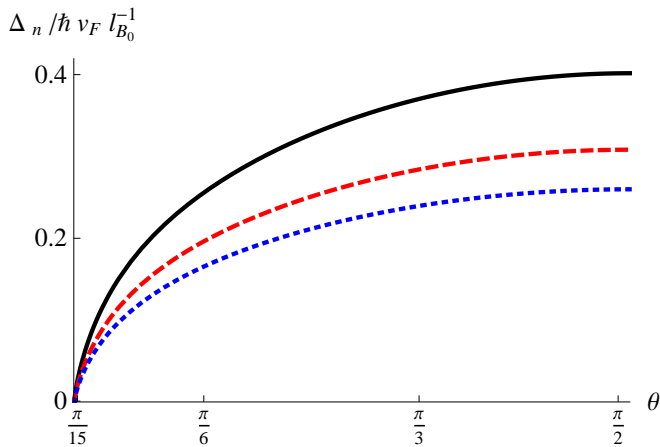}
\caption{Plot of Landau level energy gap $\De_n$ for
$n=1$ (black solid line) $2$ (red long-dashed line) and $3$ (blue
dashed line) as a function of the tilt angle $\theta$ for a fixed
electric field ${\mathcal E}=0.2 v_F B$}.\label{figseciib} \end{figure}

We also note that the magnetic field dependence of the Landau level
energy gaps are different from their counter part in graphene. To
illustrate this we define $ \De_n =E_{n+1}(k_y)-E_n(k_y)$ for $n
\neq 0$. For $\beta \ll 1 $, we find: \bea \frac{\De_n}{\hbar v_F
l_B^{0 -1}} \simeq \ga^{-3/2}\sqrt{\sin(\theta)}
\Big(\sqrt{|n+1|}-\sqrt{|n|} \Big). \label{endiff} \eea For $\beta
\simeq 1$ when $\ga \al B \gg |n|$, we find that $\De_n/(\hbar v_F
l_B^{0 -1}) \simeq \ga^{-3/2}/(2 \sqrt{\al B \ga})$. This behavior
is distinct from its counter part in graphene where $\De_n/(\hbar
v_F l_B^{0 -1})= \ga^{-3/2} \sqrt{ \sin
\theta}(\sqrt{n+1}-\sqrt{n})$ for all $B$. However, since $\al \sim
10^{-4}$/Tesla, this behavior can only be seen in a very tiny window
near critical tilt and would be hard to figure out in experiments.
On the contrary, the variation of dispersion of $\De_n$ with the
tilt angle $\theta$, seen in Fig.\ \ref{figseciib} for several $n$,
can be tested experimentally with microwave absorption experiments
routinely done for conventional quantum Hall systems \cite{qheexpt}.
This will be discussed further in Sec.\ \ref{conc}.

\begin{figure} \includegraphics[width=\linewidth]{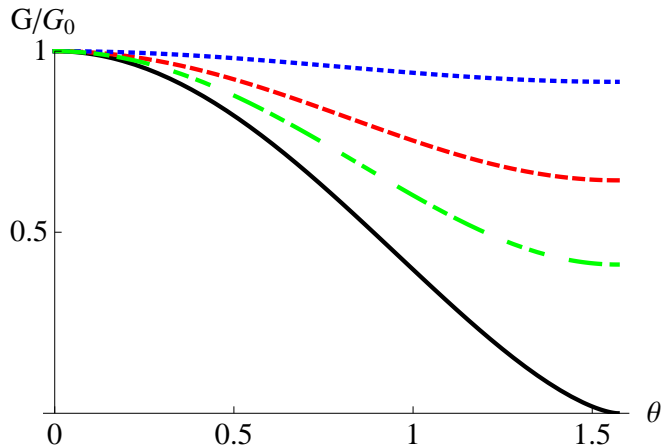}
\caption{Plot of the conductance $G/G_0$ as a function of the tilt
angle $\theta$ for electric field ${\mathcal E}=1$ (black solid
line), $1.2$ (green dash-dotted line), $1.5$ (red long dashed line),
$3$ (blue dashed line) for a fixed magnetic field $B$.}
\label{figseciic} \end{figure}

Now we turn to the solution of this problem in the regime ${\mathcal
E} \ge v_F B$ where we get scattering states. In this regime, we
define a parameter $\beta'= v_F B \sin \theta/{\mathcal E}$ and
perform a similar boost transformation as outlined earlier. This
allows us to shift to a reference frame where there is no magnetic
field and the Schr\"odinger equation, in the momentum
representation, reads \cite{levitov1} \bea i e {\mathcal E}'\hbar
\partial_{k_x} \psi' &=& \Big[ \hbar
v_F (\si_x k_x + \si_y k_y) - E' \non \\
&& - g \mu_B \si_z B \sin \theta \Big] \psi' , \label{sch2} \eea
where ${\mathcal E}'=\ga'({\mathcal E}-\beta' v_F B \sin \theta)$
and $E'= \ga'(E - \hbar v_F k_y \beta')$ are the electric field and
energy as seen in the boosted frame, and $\ga'= 1/\sqrt{1-\beta^{'
2}}$. The scattering states can now be easily obtained from this
equation by noting the similarity of this equation with the standard
Landau-Zenner problem with modified Planck's constant $\hbar \to
e{\mathcal E}' \hbar$. In particular, the transmission probability
of these Dirac electrons in the direction of the applied electric
field in the boosted frame can be written as \bea T(k_y;E) &=&
e^{-\pi d_0^2 \ga' [(g \mu_B B \sin \theta)^2 + (\hbar v_F
k_y^{'})^2]/(\hbar v_F)^2}, \label{transelec1} \eea where $d_0 =
\sqrt{\hbar v_F/(e{\mathcal E})}$ is the legth scale set by the
electric field. Now one can rotate back to the ``laboratory frame"
using $ k_y^{'} = \ga'(k_y - \beta' \ep /\hbar v_F)$, and integrate
over $k_y$ modes to obtain the tunneling conductance \cite{levitov1}
\bea G &=& G_0 (1-\beta^{'\,2})^{3/4} e^{-\pi \ga' d_0^2(g \mu_B B
\sin \theta)^2/(\hbar v_F)^2}, \eea where $G_0= e^2 L_y/(h d_0)$,
and $L_y$ is the sample width. We find that in contrast to graphene
\cite{levitov1}, the Zeeman term arising from the magnetic field
along $z$ produces an additional exponential suppression of the
conductance. This can be understood by noting that in topological
insulators, a Zeeman magnetic field along $z$ results in the
generation of a mass term for the Dirac electrons and hence leads to
a suppression of the conductance. A plot of $G/G_0$ as a function of
the tilt of a magnetic field $\theta$ is shown in Fig.\
\ref{figseciic} for several representative values of the electric
field $\mathcal E$ and for a fixed magnetic field $B$. The plot
shows that for small electric fields, the conductance is quickly
suppressed as we increase $\theta$ from $0$ to $\pi/2$; however for
larger fields, the suppression is minimal. Thus one can tune the
conductance of these insulators either by tuning the electric field
at a fixed $\theta$ or by tuning $\theta$ at a fixed electric field.

\section{Transport in NMN and NBM junctions}
\label{nmjn}

In this section, we analyze the properties of NMN and NBM junctions
of topological insulators as shown in the left and right panels of Fig.
\ref{fig2a}. Sec.\ \ref{nmn1} discusses the NMN junctions while Sec.\
\ref{nbm1} elucidates the properties of the NBM junctions.

\subsection{NMN junctions}
\label{nmn1}

The proposed experimental set up for the NMN junction is shown in
the left panel of Fig.\ \ref{fig2a}. The Dirac fermions in region I
and III are described by the Hamiltonian in Eq. (\ref{ham1}).
Consequently, the wave functions of these fermions moving along $\pm
x$ in these regions for a fixed transverse momentum $k_y$ and energy
$\ep$ can be written as
\beq \psi_j^{\pm} ~=~ (1, \pm e^{\pm i \al}) ~\exp [i (\pm k_x x + k_y y)],
\label{wav1} \eeq
where $j$ takes values I and III, and
\bea \al &=& \arcsin(\hbar v_F k_y/|\ep +\mu|), \non \\
k_x(\ep) &=& \sqrt{[(\ep+\mu)/\hbar v_F]^2 - k_y^2}. \label{wav2}
\eea In region II, the presence of the ferromagnetic strip with a
magnetization $\vec m_0 = m_0 \hat y$ leads to the additional term
\bea H_{\rm induced} = \int dx dy \, {\mathcal J} \theta(x)
\theta(d-x) \psi^\dg (\vec r) \si_y\, \psi(\vec r), \label{wav3}
\eea where ${\mathcal J} \sim m_0$ is the exchange field due to the
presence of the strip \cite{tanaka1}, and $\theta(x)$ denotes the
Heaviside step function. Note that $H_{\rm induced}$ may be thought
as a vector potential term arising due to a fictitious magnetic
field $\vec B_{f}= ({\mathcal J}/ev_F) [\delta(x)-\delta(d-x)] \hat
z$. This analogy shows that our choice of the in-pane magnetization
along $\hat y$ is completely general; all gauge invariant quantities
such as the transmission probability are independent of the
$x$-component of $\vec m_0$ in the present geometry. We emphasize
that this effect is distinct from that due to a finite $z$ component
of $\vec m_0$ which provides a mass to the Dirac electrons. For a
given $m_0$, the precise magnitude of ${\mathcal J}$ depends on the
exchange coupling of the film and can be tuned, for soft
ferromagnetic films, by an applied field \cite{tanaka1}. The wave
function for the Dirac fermions in region II moving along $\pm x$ in
the presence of such an exchange field is given by
\bea \psi_{II}^{\pm} ~=~ (1, \pm e^{\pm i \beta}) ~\exp [i( \pm k_x^{'}
x+ k_y y)], \label{wav4} \eea
where
\bea \beta &=& \arcsin(\hbar v_F (k_y + M)/|\ep +\mu|), ~~M=
{\mathcal J}/(\hbar v_F), \non \\
k_x^{'}(\ep) &=& \sqrt{[(\ep+\mu)/ \hbar v_F]^2 - (k_y+M)^2}. \label{wav5} \eea
Note that beyond a critical $M_c= \pm 2|\ep+\mu|/(\hbar v_F)$, and hence a
critical ${\mathcal J}_c = \pm 2 |\ep+\mu|$, $k^{'}_x$ becomes imaginary for
all $k_y$ leading to spatially decaying modes in region II.

Let us now consider an electron incident on region II from the left
with a transverse momentum $k_y$ and energy $\ep$. Taking into
account reflection and transmission processes at $x=0$ and $x=d$,
the wave function of the electron can be written as
\bea \psi_I &=& \psi_I^+ + r \psi_I^-, \quad \psi_{II} = p \psi_{II}^+
+ q \psi_{II}^-, \quad \psi_{III} = t \psi_{III}^+. \non \\ \label{wav6} \eea
Here $r$ and $t$ are the reflection and transmission amplitudes, and $p$ ($q$)
denotes the amplitude of right (left) moving electrons in region II. Matching
boundary conditions on $\psi_I$ and $\psi_{II}$ at $x=0$ and $\psi_{II}$ and
$\psi_{III}$ at $x=d$ leads to
\bea 1+r &=& p+q, \quad e^{i \al} -r e^{-i\al} = p e^{i \beta} - q
e^{-i \beta}, \non \\
te^{i k_x d}&=& p e^{ik_x^{'}d} + q e^{-ik_x^{'}d}, \non \\
te^{i (k_x d+\al)}&=& p e^{i(k_x^{'}d +\beta)} - q
e^{-i(k_x^{'}d+\beta)}. \label{cond1} \eea

Solving for $t$ from Eq. (\ref{cond1}), one finally obtains the conductance
\bea G = dI/dV = (G_0/2) \int_{-\pi/2}^{\pi/2} d \al ~T \,
\cos \al. \label{wav7} \eea
Here $G_0 = \rho(eV) w e^2/(\pi \hbar^2 v_F) $, $\rho(eV)= |(\mu+eV)|/ [2\pi
(\hbar v_F)^2]$ is the density of states (DOS) of the Dirac fermions and is
a constant for $\mu \gg eV$, $w$ is the sample width, and the transmission
$T = |t|^2$ is given by
\bea T &=& \cos^2(\al) \cos^2(\beta)/[\cos^2(k_x^{'} d)\cos^2(\al)
\cos^2(\beta) \non \\
&& + \sin^2 (k_x^{'} d) ( 1-\sin \al \sin \beta)^2]. \label{trans1} \eea

\begin{figure} \includegraphics[width=0.95\linewidth]{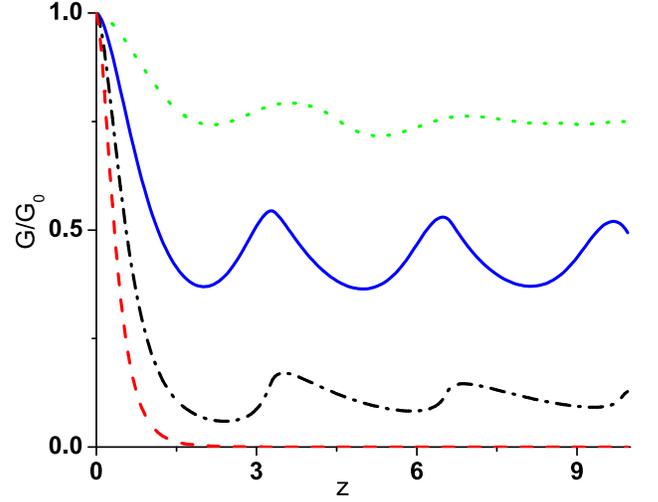}
\caption{Plot of tunneling conductance $G/G_0$ for a fixed $V$ and $\mu$ as
a function of the effective width $z=d|eV+\mu|/\hbar v_F$ for $\hbar v_F
M/|eV+\mu| =0.3$ (green dotted line), $0.7$ (blue solid line), $1.3$ (black
dash-dotted line) and $2.1$ (red dashed line). The value of the critical $M$
is given by $\hbar v_F M/|eV+\mu|=2$.} \label{fig2} \end{figure}

\begin{figure} \includegraphics[width=0.9 \linewidth]{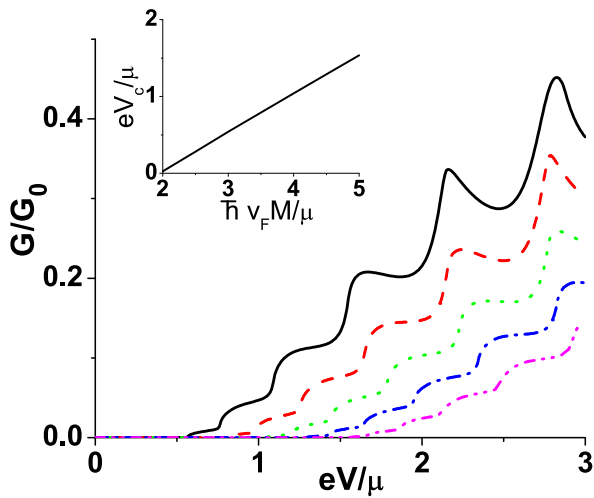}
\caption{Plot $G/G_0$ versus $eV/\mu$ for several representatives values
$\hbar v_F M/\mu$ ranging from $3$ (left-most black solid curve) to $5$
(right-most magenta dash-double dotted line) in steps of $0.5$ The effective
junction width $z_0=5$ for all plots. The inset shows a plot of $eV_c/\mu$
versus $\hbar v_F M/\mu$. See text for details.} \label{fig3} \end{figure}

Eq. (\ref{trans1}) and the
expression for $G$ represent one of the main results of this
section. We note that for a given $\al$, $T$ has an oscillatory
(monotonically decaying) dependence on $d$ provided $k_x^{'}$ is
real (imaginary). Since $k_x^{'}$ depends, for a given $\al$, on
$M$, we find that one can switch from an oscillatory to a
monotonically decaying $d$ dependence of transmission in a given
channel (labeled by $k_y$ or equivalently, $\al$) by turning on a
magnetic field which controls $m_0$ and hence $M$. Also, since $-1
\le \sin \al \le 1$, we find that beyond a critical $M=M_c$, the
transmission in all the channels exhibits a monotonically decaying
dependence on $d$. Consequently, for a thick enough junction one can
tune $G$ at fixed $V$ and $\mu$ from a finite value to nearly zero
by tuning $M$ ({\it i.e.}, $m_0$) through $M_c$. Thus such a
junction may be used as a magnetic switch. These qualitatively
different behaviors of the junction conductance $G$ for $M$ below
and above $M_c$ is demonstrated in Fig.\ \ref{fig2} by plotting $G$
as a function of the effective barrier width $z=d|eV+\mu|/\hbar v_F$
for several representative values of $\hbar v_F M/|eV+\mu|$. Since
$T$ and hence $G$ depends on $M$ through the dimensionless parameter
$\hbar v_F M/|eV+\mu|$, this effect can also be observed by varying
the applied voltage $V$ for a fixed $\mu$, $d$, and $M$. In that
case, for a reasonably large dimensionless barrier thickness $z_0=d
\mu/\hbar v_F$, $G/G_0$ becomes finite only beyond a critical
voltage $|eV_c + \mu| = \hbar v_F M/2$ as shown in Fig.\ \ref{fig3}
for several representative values of $z_0$. The critical voltage
$V_c$ can be determined numerically by finding the lowest voltage
for which $G/G_0$ exhibits a monotonic decay as a function of $z_0$.
The plot of $eV_c/\mu$ as a function of $\hbar v_F M/\mu$, shown in
the inset of Fig.\ \ref{fig3}, demonstrates the expected linear
relationship between $V_c$ and $M$. We note that such a dependence
of $G$ on $M$ or $V$ requires the Dirac electrons to be spinors in
physical spin space, and is therefore impossible to achieve in
either graphene \cite{neto1} or in a conventional 2D electron gas
for which a proximate ferromagnetic film would only provide a Zeeman
term for the electrons, leaving $G$ unaffected.

\subsection{NBM junctions}
\label{nbm1}

Next, we analyze the NBM junction shown in the right panel of Fig.\
\ref{fig2a} where the region III below a ferromagnetic film is
separated from region I by a potential barrier in region II. Such a
barrier can be applied by changing the chemical region in region II
either by a gate voltage $V_0$ or via doping \cite{exp2}.
We will analyze the problem in the thin barrier
limit in which $V_0 \to \infty$ and $d\to 0$, keeping the
dimensionless barrier strength $\chi= e V_0 d/(\hbar v_F)$ finite.
The wave function of the Dirac fermions moving along $\pm x$ with a
fixed momentum $k_y$ and energy $\ep$ in this region is given by
\bea \psi^{'\,\pm}_{II} ~=~ (1, \pm e^{\pm i \ga}) ~\exp [i
(\pm k^{''}_x x + k_y y)/\sqrt{2}], \label{wav8} \eea
where
\bea \ga &=& \arcsin( \hbar v_F k_y/|\ep + eV_0+\mu|), \non \\
k^{''}_x (\ep) &=& \sqrt{[(\ep+eV_0+\mu)/\hbar v_F]^2 -
k_y^2}. \label{wav9} \eea
The wave functions in region I and III are given by
$\psi^{'}_I = \psi_I$ and $\psi^{'}_{III} = \psi_{II}$, where $\psi_I$
and $\psi_{II}$ are given in Eq. (\ref{wav6}). Note that one
can have a propagating solution in region III only if $|M| \le |M_c|$.

\begin{figure} \includegraphics[width=0.95 \linewidth]{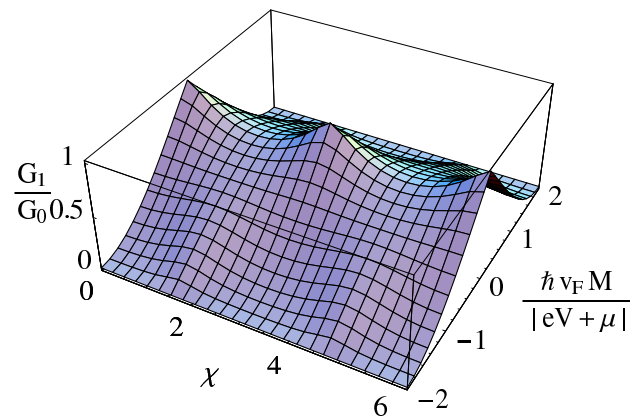}
\caption{Plot of tunneling conductance $G_1/G_0$ versus $\chi$ and
$M$ for a fixed applied voltage $V$ and chemical potential $\mu$. $G_1$
vanishes for $|M| \ge M_c=2|eV+\mu|/\hbar v_F$.} \label{fig4} \end{figure}

The transmission problem for such a junction can be solved by a
procedure similar to the one outlined above for the magnetic strip
problem. For an electron approaching the barrier region from the left,
we write down the following forms of the wave function in the three
regions I, II and III: $\psi^{'}_I = \psi_I^+ + r_1 \psi_I^-$,
$\psi^{'}_{II} = p_1 \psi^{'\,+}_{II} + q_1 \psi^{'\,-}_{II}$, and
$\psi^{'}_{III} = t_1 \psi_{II}^+$. As outlined earlier, one can
then match boundary conditions at $x=0$ and $x=d$, and obtain the
transmission coefficient $T_1=|t_1|^2 k^{'}_x/k_x$ as
\bea T_1 &=& 2 \cos \beta \cos \al/[1+\cos(\beta-\al) \non \\
&& -\cos^2(\chi)\{\cos(\beta -\al)-\cos(\beta +\al)\}]. \label{trans2} \eea
Note that in the absence of the ferromagnetic
film over region III, $\beta=\al$, and $T_1 \to T^0_1 =
\cos^2(\al)/[1-\cos^2(\chi) \sin^2 (\al)]$. The expression for
$T_1^0$, reproduced here for the special case of $M=0$, is well
known from analogous studies in the context of graphene, and it
exhibits both the Klein paradox ($T^0_1=1$ for $\al=0$) and
transmission resonances ($T^0_1=1$ for $\chi=n \pi $) \cite{kat1}.
When $M\ne 0$, we find that the transmission for normal incidence
($k_y=0$) does become independent of the barrier strength, but its
magnitude deviates from unity:
\beq T^{\rm normal}_1 = \frac{2\sqrt{1-(\hbar
v_F M/|eV +\mu|)^2}}{1+ \sqrt{1-(\hbar v_F M/|eV +\mu|)^2}}. \eeq
The value of $T^{\rm normal}_1$ decreases monotonically from $1$ for
$M=0$ to $0$ for $|M| = |eV +\mu|/(\hbar v_F)$, and can thus be tuned
by changing $M$ (or $V$) for a fixed $V$ (or $M$) and $\mu$.

\begin{figure} \includegraphics[width=0.9 \linewidth]{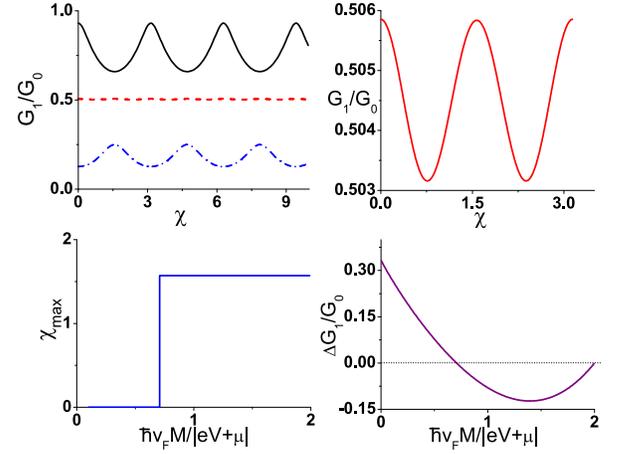}
\caption{Top left panel: Plot of $G_1/G_0$ versus $\chi$ for $\hbar
v_F M/|eV+\mu|=0.1$ (black solid line), $0.7075$ (red dashed line), and $1.4$
(blue dash-dotted line) for fixed $V$ and $\mu$. Top right panel: Plot of
$G_1/G_0$ versus $\chi$ at $M=M^{\ast}$ showing the period halving. Bottom
left panel: Plot of $\chi_{\rm max}$ versus $\hbar v_F M/|eV+\mu|$ showing
conductance maxima positions. Bottom right panel: Plot of $\De G_1/G_0$
versus $\hbar v_F M/|eV+\mu|$ which crosses $0$ at $M= M^{\ast}$. The dotted
line is a guide to the eye.} \label{fig5} \end{figure}

The conductance of this junction is given by $G_1 = (G_0/2)
\int_{-\al_1}^{\al_2} d \al T_1 \cos \al$, where
$\al_{1,2}$ are determined from the solution of $\cos \beta=0$
for a given $M$. A plot of $G_1$ as a function of $\hbar v_F
M/|eV+\mu|$ and $\chi$ (for a fixed $eV$ and $\mu$) is shown in
Fig.\ \ref{fig4}. We find that the amplitude of $G_1$ decreases
monotonically as a function of $|M|$ reaching $0$ at $M=M_c$ beyond
which there are no propagating modes in region III. Also, as we
increase $M$, the conductance maxima shifts from $\chi= n \pi$ to $
\chi= (n+1/2)\pi$ beyond a fixed value of $M^{\ast}(V) \simeq \pm
c_0 |eV+\mu|/(\hbar v_F)$ as shown in the top left panel of Fig.\
\ref{fig5}. Numerically, we find $c_0=0.7075$. At $M=M^{\ast}$,
$G_1(\chi=n\pi)=G_1(\chi=(n+1/2)\pi)$, leading to a period halving
of $G_1(\chi)$ from $\pi$ to $\pi/2$ . This is shown in the top right
panel of Fig.\ \ref{fig5} where $G_1(M=M^{\ast})$ is plotted as a
function of $\chi$. We note that near $M^{\ast}$, the amplitude of
oscillation of $G_1$ as a function of $\chi$ becomes very small so
that $G_1$ is almost independent of $\chi$. In the bottom left panel
of Fig.\ \ref{fig5}, we plot $\chi=\chi_{\rm max}$ (the value of
$\chi$ at which the first conductance maxima occurs) as a function
of $\hbar v_F M/|eV+\mu|$ which clearly demonstrates the shift. This
is further highlighted by plotting $\De G_1 =G_1(\chi=0)-G_1(\chi=\pi/2)$
as a function of $\hbar v_F M/|eV+\mu|$ in the bottom right panel of
Fig.\ \ref{fig5}. $\De G_1$ crosses zero at $M=M^{\ast} <M_c$ indicating the
position of the period halving. Thus the position of the conductance maxima
depends crucially on $\hbar v_F M/(eV+\mu)$ and can be tuned by
changing either $M$ or $V$.

\subsection{Alternative forms of the Hamiltonian}
\label{twoforms}

In this subsection, we discuss a possible way of distinguishing
between possible forms of the Dirac Hamiltonian in the surface of a
topological insulator. In the literature (see, for instance, Ref.
\onlinecite{xu}), two such different forms have been studied for the
first part of the Hamiltonian in Eq. (\ref{ham1}), namely,
\bea h_1 &=& \int \frac{d^2k}{(2 \pi)^2} \psi^\dg (\vec k) ~\hbar v_F (\si_x
k_x ~+~ \si_y k_y ) ~\psi (\vec k) \non \\
{\rm and} ~~h_2 &=& \int \frac{d^2k}{(2 \pi)^2} \psi^\dg (\vec k)
~\hbar v_F (\si_x k_y ~-~ \si_y k_x) ~\psi (\vec k). \label{ham2}
\nonumber\\ \eea We have implicitly assumed the form $h_1$ in the
entire analysis in this paper. We note that $h_1$ and $h_2$ are both
time-reversal invariant since ${\vec \si} \to - {\vec \si}$ and
${\vec k} \to - {\vec k}$ under that transformation, and they are
also invariant under rotations in the $x - y$ plane. But under the
two-dimensional parity transformation $x \to - x$ and $y \to y$,
they transform differently; since $k_x \to - k_x$, $k_y \to k_y$,
$\si_x \to \si_x$ and $\si_y \to - \si_y$, we see that $h_1 \to -
h_1$ while $h_2 \to h_2$. Since the Hamiltonian of the surface Dirac
electrons arises from a spin-orbit coupling in the bulk which is
then projected on to the two-dimensional surface, and we have not
discussed the bulk Hamiltonian here, we have no {\it a priori}
reason to choose between $h_1$ and $h_2$. In principal, we could
even consider a linear combination of the two such as $\cos \theta
~h_1 + \sin \theta ~h_2$. Clearly, when an in-plane magnetization
which breaks the in-plane rotational symmetry is introduced using
the ferromagnetic film, the effect of this on the analysis in Secs.
\ref{nmn1} and \ref{nbm1} will depend on the angle $\theta$
mentioned above; for instance, a magnetization in the $y$ direction
will couple to $\si_y$ and will therefore shift the momentum $k_y$
for $h_1$ and $k_x$ for $h_2$. Hence, when experimental tests of the
various results obtained in those two sections are performed, one
can probe whether the Hamiltonian for the system of interest is
actually $h_1$ or $h_2$ or a linear combination of the two, by
varying the direction of magnetization of the ferromagnetic film and
studying the effect that this has on the conductance. For example,
if the hamiltonian describing the surface electrons of the
topological insulator turns out to be $h_2$, $m_y$ will have no
effect on transport. In general, for any $\theta$, there will be
specific direction of the in-plane magnetization $\vec m \equiv m_y
\cos \theta +m_x \sin \theta$ which will have maximal effect on the
transport while the component of the magnetization $ \vec m' = - m_x
\cos \theta+ m_y \sin \theta$ will not affect the transport at all.

\section{Transport in NMS junctions}
\label{supjn}

We consider a NMS junction on the surface of a topological insulator
as shown in Fig.\ \ref{fig3a}. As shown there, region
II, which extends from $x=-d$ to $x=0$, has a proximate
ferromagnetic film leading to an induced magnetization $\vec M = M
\hat y$. Region III depicts the superconducting region occupying
$x>d$. We assume that superconductivity in this regime is induced
via a proximate superconducting film with $s$-wave pairing as shown
in the figure. The quasiparticles of such a superconductor can be described
by the following Dirac-Bogoliubov-de Gennes equation \cite{kane2}
\beq \left( \begin{array}{cc}
H-E_F & \De (r) \\
\De^*(r) & E_F-H \end{array} \right) \psi=E\psi, \eeq
where $\psi=( \psi_\uparrow, \psi_\downarrow,
\psi^\dg_\uparrow, \psi^\dg_\downarrow)$ are the four components for the
electron and the hole spinors, and the Hamiltonian $H$ is given by
\beq H=-i\hbar v_F {\vec \si} \cdot {\vec \nabla} +\mu \si_y B \theta(x+d)
\theta(-x), \eeq
where $\theta(x+d)$ and $\theta(-x)$ are Heaviside step functions. $\De (r)
=\De_0 \exp(i\phi)\theta(x)$ is the BCS pair-potential in region III.

Eq. (1) can be solved for the normal, magnetic and superconducting
regions. In the normal region, the wave functions for electron and hole
moving in $\pm x$ direction are given by
\bea \psi_N^{e\pm} &=& ( 1, \pm e^{\pm i\al}, 0, 0) \exp[i(\pm k_n x+qy)], \\
\psi_N^{h\pm} &=& ( 0, 0, 1, \mp e^{\pm i\al^{'}}) \exp[i(\pm k_n^{'} x+qy)],\\
\sin \al &=&\frac{\hbar v_F q}{|\ep+E_F|},\hspace{0.2cm} \sin
\al^{'} =\frac{\hbar v_F q}{|\ep-E_F|}, \label{nmseq1} \eea
where the wave vector $k_n(k_n^{'})$ for the electron (hole) wave functions
are given by
\beq k_n(k_n^{'})=\sqrt{\left( \frac{\ep+(-)E_F}{\hbar v_F}\right)^2-q^2},
\eeq
and $\al(\al^{'})$ is the angle of incidence of the electron (hole).

In region II, the wave functions for an electron and a hole moving in the $\pm
x$ direction are as follows:
\bea \psi_B^{e\pm} &=& ( 1, \pm e^{\pm i\theta}, 0, 0)\exp[i(\pm k_b x+qy)], \\
\psi_B^{h\pm} &=& (0,0,1,\mp e^{\pm i\theta^{'}}) \exp[i(\pm k_b^{'} x+qy)], \\
\sin \theta &=&\frac{\hbar v_F (q+M)}{|\ep+E_F|},\hspace{0.2cm} \sin
\theta^{'} =\frac{\hbar v_F (q+M)}{|\ep-E_F|}, \label{nmseq2} \eea
where the wave vector $k_b(k_b^{'})$ of the electron (hole) wave
function is given by
\beq k_b(k_b^{'})=\sqrt{\left( \frac{\ep+(-)E_F}{\hbar v_F}\right)^2-(q+M)^2}.
\label{kbeq} \eeq
Here $\theta(\theta^{'})$ is the angle of incidence of the electron
(hole). Note that in principle, we could have applied an additional
gate voltage $V_0$ in this region as was done in Ref.\ \onlinecite{kat1}.
However, this leads to an expression of the longitudinal momentum
\beq k_b(k_b^{'})=\sqrt{\left( \frac{\ep+(-)(E_F-V_0)}{\hbar v_F}\right)^2-
(q+M)^2}. \eeq
This shows that in the limit of large $V_0$, the effect of $M$ on
$k_b(k_b')$ and hence on $G$ becomes negligible. Therefore we
restrict ourselves to the $V_0=0$ limit.

In the superconducting region, the BdG quasiparticles are mixtures of
electron and holes. Hence the wave function for BdG quasiparticles
moving in $\pm x$ directions with transverse momenta $q$ and energy
$\ep$ for $E_F\gg \ep,\De_0$ are given by
\bea \psi_S^{e\pm} &=& (e^{\mp i\beta}, \mp e^{\pm i(\ga-\beta)}, e^{-i\phi},
\mp e^{i(\pm \ga -\phi)}) \non \\
&& \times \exp[i(\pm k_s x+qy)-\kappa x], \\
\sin \ga &=&\frac{\hbar v_F q}{|E_F|},\quad k_s=\sqrt{\left(
\frac{E_F}{\hbar v_F}\right)^2-q^2}, \label{nmseq3} \eea
and $\beta=\cos^{-1}(\ep / \De_0)\theta (\De_0-\ep)-i\cosh^{-1}(\ep /\De_0)
\theta(\ep-\De_0)$ where $\theta$ denotes the Heaviside step function.

Next, we note that for any transmission process to take place we
need $\al^{'},\theta ,\theta^{'} ,\ga \le \pi/2$. This condition gives
the limits for the range of $\al$. For simplicity we consider
$V_0=0$ in region II. Then $\theta^{'}>\theta>\al^{'}$. Using Eqs.
(\ref{nmseq1}), (\ref{nmseq2}) and (\ref{nmseq3}), we find that
the Andreev process takes place for $\al_{c1}<\al<\al_{c2}$, where
\bea \al_{c1} &=& \arcsin[(-|\ep -E_F|)/|\ep+E_F|], \\
\al_{c2} &=& \arcsin[(|\ep -E_F|-M)/|\ep+E_F|]. \eea
Note that $\al_{c1} \ne -\al_{c2}$, and this asymmetry is
generated by the induced magnetization $M$.

Following Ref.\ \onlinecite{kat1}, we write wave functions for the
normal, magnetic and superconducting regions as
\bea \Psi_N &=& \psi_N^{e+}+r\psi_N^{e-}+r_A\psi_N^{h-}, \\
\Psi_B &=& p\psi_B^{e+}+q\psi_B^{e-}+m\psi_B^{h+}+n\psi_B^{h-}, \\
\Psi_S &=& t\psi_S^+ +t^{'}\psi_S^- , \eea
where both normal and Andreev reflection are taken into account. Here $r$
and $r_A$ denote the amplitudes for normal and Andreev reflection
respectively. These wave functions must satisfy the following boundary
conditions,
\beq \Psi_N|_{x=-d}=\Psi_B|_{x=-d}, \hspace{0.4cm} \Psi_B|_{x=0}=\Psi_S|_{x=0}.
\eeq

\begin{figure} \includegraphics[width=0.9 \linewidth]{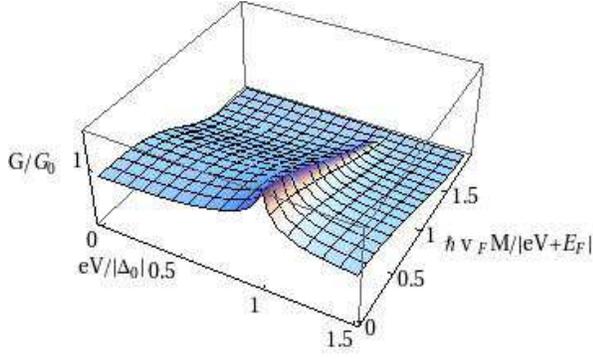}
\caption{Plot of subgap tunneling conductance $G(V)$ as a function of the
induced magnetization $M$ and applied voltage $V$. See text for details.}
\label{fignms1} \end{figure}

Solving these boundary conditions, we obtain for $r,r_A,t$ and
$t^{'}$ \cite{kat1}
\bea r &=& e^{-2ik_nd} N/D, \\
N &=& [e^{i\al}\cos(k_bd+\theta)+ i\sin(k_bd)] \non \\
&& -\rho[\cos(k_bd-\theta) + ie^{i\al}\sin(k_bd)], \\
D &=& [e^{-i\al}\cos(k_bd+\theta)-i\sin(k_bd)] \non \\
&& +\rho[\cos(k_bd-\theta)-ie^{-i\al}\sin(k_bd)], \\
t^{'} &=& \frac{1}{\cos(\theta)[\Gamma e^{-i\beta}+e^{i\beta}]}
\big(e^{-i k_n d} [\cos(k_bd-\theta)\non \\
&& +ie^{i\al}\sin(k_bd)] +re^{ik_nd}[\cos(k_bd \non \\
&& -\theta)-ie^{-i\al}\sin(k_bd)]\big), \\
t &=& \Gamma t^{'}, \\
r_A &=& \frac{t^{'}(\Gamma +1)e^{ik_n^{'}d}\cos(\theta^{'})e^{-i\phi}}{\cos
(k_b^{'}d-\theta^{'})-ie^{-i\al^{'}}\sin(k_b^{'}d)}, \eea
where the parameters $\rho$, $\Gamma$ and $\eta$ can be expressed as
\bea \rho &=& \frac{-\Gamma e^{i(\ga -\beta)}+e^{-i(\ga -\beta)}}{\Gamma
e^{-i\beta}+e^{i\beta}}, \\
\Gamma &=& \frac{e^{-i\ga}-\eta}{e^{i\ga}+\eta}, \\
\eta &=& \frac{e^{-i\al^{'}}\cos(k_b^{'}d+\theta^{'})-i\sin(k_b^{'}d)}{\cos
(k_b^{'} d-\theta^{'})-ie^{-i\al^{'}}\sin(k_b^{'}d)}. \eea

The tunneling conductance of the NMS junction can be expressed in
terms of $r$ and $r_A$ as
\bea \frac{G(eV)}{G_0(eV)}=\int_{\al_{c1}}^{\al_{c2}} d \al \left( 1-|r|^2 +
|r_A|^2\frac{\cos \al^{'}}{\cos \al} \right)\cos \al. \non \\ \eea

\begin{figure} \includegraphics[width=0.9 \linewidth]{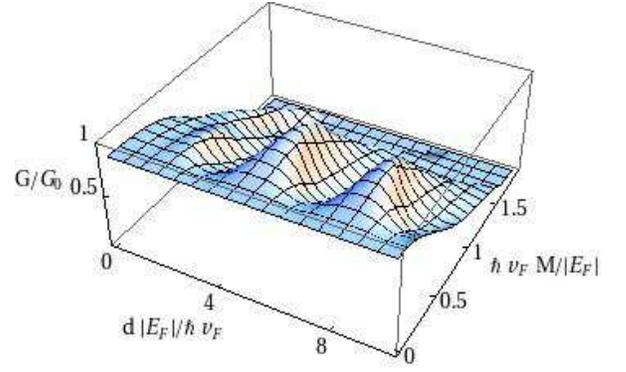}
\caption{Plot of zero-bias tunneling conductance $G(0)/G_0$ as a
function of the induced magnetization $M$ and the barrier width $d$.
See text for details.} \label{fignms2} \end{figure}

A plot of the subgap tunneling conductance $G/G_0$ as a function of
the magnetization $M$ and the applied voltage $V$ for a fixed
barrier width $d$ is shown in Fig.\ \ref{fignms1}. We find that
$G(0)$ decreases monotonically as a function of the magnetization
for all values of the applied voltage. This can be easily attributed
to a decrease in the number of conduction channels ({\it i.e.},
number of $k_y$ modes with real $k_b$) with increasing $M$. The
behavior of the zero-bias conductance as a function of the barrier
width $d$ and magnetization $M$ is shown in Fig.\ \ref{fignms2}. We
find that the zero-bias conductance shows an oscillatory behavior as
a function of the barrier width $d$ for small $M$ \cite{bhatt1}.
With increasing $M$, the position of the conductance maxima shifts
which demonstrates the tunability of the zero-bias conductance with
the induced magnetization. This continuous shift in position of the
zero-bias conductance maxima is to be contrasted with the sudden
change of its counterpart in NMN junctions of topological
insulators.

\section{Experiments}
\label{conc}

Experimental verification of our work would involve carrying out the
following experiments. For a topological insulator in a crossed
electric and magnetic field with ${\mathcal E} \le v_F B
\sin(\theta) $, we propose measurement of the energy gap of the
Landau levels as a function of the electric field strength and the
tilt angle $\theta$. Such measurements have been done in quantum
Hall systems using microwave absorption techniques \cite{qheexpt}.
The variation of the excitation energy gap between the ground and
the first excited states, $\De_1$, as shown in Fig.\
\ref{figseciib}, should be observable in similar experiments
performed with topological insulators. For ${\mathcal E} \ge v_F B
\sin(\theta)$, we propose measurement of conductance $G$ of these
films as a function of both the electric field ${\mathcal E}$ and
the tilt angle $\theta$. We predict that for small ${\mathcal E}$,
the tunneling should show a faster suppression with increase
$\theta$ from $0$ to $\pi/2$. We note that for a $\theta$ suitably
chosen between $0$ and $\pi/2$, the conductance of these films can
tuned via an electric field, as demonstrated in Fig.\
\ref{figseciic}, leading to realization of electric field controlled
switching in these materials. For the NMN and NBM junctions, which
can be prepared by depositing ferromagnetic films on the surface of
a topological insulator, we propose measurement of the tunneling
conductance $G$ as a function of $m_0$. For the geometry shown in
the left panel of Fig.\ \ref{fig2a}, we predict that depending on
the magnetization $M$, $G$ should demonstrate either a monotonically
decreasing or an oscillatory behavior as a function of the junction
width $d$. Another, probably more experimentally convenient, way to
realize this effect would be to measure $V_c$ of a junction of width
$d$ for several values of $M$ and confirm that $V_c$ varies linearly
with $M$ with a slope of $\hbar v_F/(2e)$, provided $\mu$ and $d$
remain fixed. For the geometry depicted in the right panel of Fig.\
\ref{fig2a}, one would, in addition, need to create a barrier by
tuning the chemical potential of an intermediate thin region of the
sample as done earlier for graphene \cite{neto1}. Here we also
propose measurement of $G_1$ as a function of $V_0$ (or equivalently
$\chi$) for several representative values of $m_0$ and a fixed $V$.
We predict that the maxima of the tunneling conductance would shift
from $\chi=n \pi$ to $\chi=(n+1/2)\pi$ beyond a critical $m_0$ for a
fixed $V$, or equivalently, below a critical $V$, for a fixed $m_0$.
Finally, for the NMS junction, we propose measurement of the
tunneling conductance $G(V)$ as a function of the magnetization $M$
which should demonstrate the decaying behavior shown in Fig.\
\ref{fignms1}. The tunability of the zero-bias conductance maxima,
shown in Fig.\ \ref{fignms2}, can also be tested by making junctions
with different widths.

In conclusion, we have studied several magnetotransport properties
of Dirac Fermions on the surface of a topological insulator, and have
shown that they exhibit several properties which are distinct both
from their counterparts in graphene and conventional Schrodinger
electrons in other 2D systems. These novel features include tunability
of the orbital and Zeeman effects of an applied magnetic field
with a crossed in-plane electric field, realization of a magnetic
switch using a NMN junction, and magnetic tunability of transmission
resonances of Dirac fermions in NBM and NMS junctions. We have
suggested experiments which can verify our theory.

\section*{Acknowledgments}

D.S. thanks DST, India for financial support under Project No.
SR/S2/CMP-27/2006. K.S. thanks DST, India for financial support
under Project No. SR/S2/CMP-001/2009 and K. Ray for several
illuminating discussions on related topics.


\begin{thebibliography}{99}

\bibitem{zhang1} B. A. Bernevig, T. L. Hughes, and S.-C. Zhang, Science
{\bf 314}, 1757 (2006); B. A. Bernevig and S.-C. Zhang, Phys. Rev. Lett.
{\bf 96}, 106802 (2006).

\bibitem{hasan1} M. K\"onig, S. Wiedmann, C. Br\"une, A. Roth, H. Buhmann,
L. W. Molenkamp, X.-L. Qi, and S.-C. Zhang, Science {\bf 318}, 766 (2007);
D. Hsieh, D. Qian, L. Wray, Y. Xia, Y. S. Hor, R. J. Cava, and M. Z. Hasan,
Nature {\bf 452}, 970 (2008).

\bibitem{kane1} C. L. Kane and E. J. Mele, \prl{\bf 95}, 226801 (2005);
{\it ibid}, \prl{\bf 95}, 146802 (2005).

\bibitem{kane2} L. Fu, C. L. Kane, and E. J. Mele, \prl{\bf 98}, 106803 (2007);
R. Roy, \prb{\bf 79}, 195322 (2009); J. E. Moore and L. Balents, \prb{\bf 75},
121306(R) (2007).

\bibitem{qi1} X. L. Qi, T. L. Hughes, and S. C. Zhang, Phys. Rev. B {\bf 78},
195424 (2008); H. Zhang, C.-X. Liu, X.-L. Qi, X. Dai, Z. Fang, and S.-C.
Zhang, Nature Phys. {\bf 5}, 438 (2009).

\bibitem{exp2} Y. Xia, D. Qian, D. Hsieh, L. Wray, A. Pal, H. Lin, A. Bansil,
D. Grauer, Y. S. Hor, R. J. Cava, and M. Z. Hasan, Nature Phys. {\bf 5}, 398
(2009); Y. Xia, D. Qian, D. Hsieh, R. Shankar, H. Lin, A. Bansil, A. V.
Fedorov, D. Grauer, Y. S. Hor, R. J. Cava and M.Z. Hasan, arXiv:0907.3089
(unpublished).

\bibitem{expt1} Y. L. Chen, J. G. Analytis, J.-H. Chu, Z. K. Liu, S.-K. Mo,
X. L. Qi, H. J. Zhang, D. H. Lu, X. Dai, Z. Fang, S. C. Zhang, I. R. Fisher,
Z. Hussain, and Z.-X. Shen, Science {\bf 325}, 178 (2009); T. Zhang,
P. Cheng, X. Chen, J.-F. Jia, X. Ma, K. He, L. Wang, H. Zhang, X. Dai, Z.
Fang, X. Xie, Q.-K. Xue, arXiv:0908.4136v3 (unpublished).

\bibitem{hasan2} D. Hsieh, Y. Xia, D. Qian, L. Wray, J. H. Dil, F. Meier,
J. Osterwalder, L. Patthey, J. G. Checkelsky, N. P. Ong, A. V. Fedorov, H.
Lin, A. Bansil, D. Grauer, Y. S. Hor, R. J. Cava, and M. Z. Hasan, Nature
{\bf 460}, 1101 (2009); P. Roushan, J. Seo, C. V. Parker, Y. S. Hor, D. Hsieh,
D. Qian, A. Richardella, M. Z. Hasan, R. J. Cava, and A. Yazdani, Nature
{\bf 460}, 1106 (2009); D. Hsieh, Y. Xia, L. Wray, D. Qian, A. Pal, J. H. Dil,
J. Osterwalder, F. Meier, G. Bihlmayer, C. L. Kane, Y. S. Hor, R. J. Cava,
and M. Z. Hasan, Science {\bf 323}, 919 (2009).

\bibitem{kane4} L. Fu and C. L. Kane, \prl{\bf 100}, 096407 (2008).

\bibitem{been1} A. R. Akhmerov, J. Nilsson, and C. W. J. Beenakker, \prl{\bf
102}, 216404 (2009); Y. Tanaka, T. Yokoyama, and N. Nagaosa, \prl{\bf 103},
107002 (2009).

\bibitem{tanaka1} T. Yokoyama, Y. Tanaka, and N. Nagaosa, arXiv:0907.2810v2
(unpublished).

\bibitem{mondal1} S. Mondal, D. Sen, K. Sengupta, and R. Shankar, \prl{\bf
104}, 046403 (2010).

\bibitem{neto1} A. H. Castro Neto, F. Guinea, N. M. R. Peres, K. S. Novoselov,
and A. K. Geim, Rev. Mod. Phys. {\bf 81}, 109 (2009); C. W. J. Beenakker,
Rev. Mod. Phys. {\bf 80}, 1337 (2008).

\bibitem{kat1} M. I. Katsnelson, K. S. Novoselov, and A. K. Geim, Nature
Phys. {\bf 2}, 620 (2006); C. W. J. Beenakker, \prl{\bf 97}, 067007
(2006); S. Bhattacharjee and K. Sengupta, \prl{\bf 97}, 217001 (2006).

\bibitem{comment1} Analogous situations may arise for graphene electrons in
the presence of suitable gate voltages, but not ferromagnetic films. See M. M.
Fogler, F. Guinea, and M. I. Katsnelson, \prl{\bf 101}, 226804 (2008).

\bibitem{comment2} The orbital effect of a magnetic field along $z$ in a
multiple-barrier geometry may also reduce transmission in single and bilayer
graphene. However, this property does not rely on the Dirac nature of graphene
electrons. See M. R. Masir, P. Vasilopoulos, and F. M. Peeters, App. Phys.
Lett. {\bf 93}, 242103 (2008).

\bibitem{vinu1} V. Lukose, R. Shankar, and G. Baskaran, \prl{\bf 98}, 116802
(2007).

\bibitem{qheexpt} A. Pinczuk, B. S. Dennis, L. N. Pfeiffer, and K. West,
Phys. Rev. Lett. {\bf 70}, 3983 (1993).

\bibitem{levitov1} A. Shytov, N. Gu, and L. Levitov, arXiv:0708.3081
(unpublished); A. Shytov, M. Rudner, and L. Levitov \prl{\bf 101}, 156804
(2008).

\bibitem{xu} C. Xu, arXiv:0909.2647v3 (unpublished).

\bibitem{bhatt1} S. Bhattacharjee, M. Maiti, and K. Sengupta, \prb{\bf 76},
184514 (2007).

\end{thebibliography}
\end{document}